\documentclass[12pt]{article}
\usepackage{subeqn}
\usepackage{epsfig}
\newcommand{\be}{\begin{equation}}
\newcommand{\ee}{\end{equation}}

\newcommand{\no}{\noindent}
\newcommand{\ce}{\begin{center}}
\newcommand{\nc}{\end{center}}

\makeatletter
\@addtoreset{equation}{section}
\makeatother

\def\sqr#1#2{{\vcenter{\vbox{\hrule height.#2pt
\hbox{\vrule width.#2pt height#1pt \kern#1pt
\vrule width.#2pt} \hrule height.#2pt}}}}

\def\operp{\hbox{${\kern+.25em{\bigcirc}
\kern-.85em\bot\kern+.85em\kern-.25em}$}}

\def\lsim{\;\raise0.3ex\hbox{$<$\kern-0.75em\raise-1.1ex\hbox{$\sim$}}\;}
\def\gsim{\;\raise0.3ex\hbox{$>$\kern-0.75em\raise-1.1ex\hbox{$\sim$}}\;}

\def\no{\noindent}

\def\ce{\centerline}
\def\ve{\vfill\eject}
\def\rdots{\mathinner{\mkern1mu\raise1pt\vbox{\kern7pt\hbox{.}}\mkern2mu
\raise4pt\hbox{.}\mkern2mu\raise7pt\hbox{.}\mkern1mu}}

\def\e e{$e^+ e^-$ }

\begin{document}

\ce{\bf THE STRONG AND GRAVITATIONAL COUPLINGS} 
\ce{\bf OF KNOTTED SOLITONS}

\vskip.3cm

\ce{\it Robert J. Finkelstein}
\vskip.3cm

\ce{Department of Physics and Astronomy}
\ce{University of California, Los Angeles, CA 90095-1547}

\vskip1.0cm

\no {\bf Abstract.}  We extend our earlier study of the electroweak
interactions of quantum knots to their gravitational and strong interactions.
The knots are defined by appropriate quantum groups and are intended to
describe all knotted field structures that conserve mass and spin, charge
and hypercharge, as well as color charge and color hypercharge.  As sources
of the gravitational fields the knots are described as representations of 
the quantum group $SL_q(2)$ and as sources of the electroweak and strong
fields they are described by $SU_q(2)$.  When the point sources of the
standard theory are replaced by the quantum knots, the interaction terms of
the new Lagrangian density acquire knot form factors and the standard local
gauge invariance is supplemented by an additional global $U(1)\times U(1)$
invariance of the $SU_q(2)$ algebra.

\ve

\section{Introduction.}

In this paper we are interested in extending our previous discussion of
knots as sources of the electroweak interactions$^{1,2,3}$ to their
gravitational and strong couplings.  In the standard theory the elementary
fermions are sources of the weak, strong, and gravitational fields.  They
are labelled by quantum numbers originating in the corresponding symmetry
groups $SU(2)\times U(1)$, $SU(3)$, and Lorentz respectively, and the
connections of these groups give rise through the quantum theory of fields
to electroweak bosons, gluons, and gravitons, respectively.  The fermionic
sources are usually understood to be structureless point particles, but
this assumption cannot be correct if one believes that some structure is
needed to endow the sources with their quantum numbers.  We shall
hypothesize that the sources are solitons and further that the solitons
are quantum knots.

All knots, either oriented or unoriented, may be described by
$SL_q(2)$.$^4$  We shall describe the subset of oriented knots not only
by $SL_q(2)$ but also by the subgroup $SU_q(2)$.  We shall also assume that
all knots, either oriented or unoriented, are gravitationally coupled but
that only the oriented knots are coupled to the electroweak and gluon fields.  Under
this assumption the non-oriented knots are candidates for dark matter.
In describing the gravitational interactions we shall assume that the
internal state functions of all knots appear as irreducible representations
of $SL_q(2)$ while in the case of the electroweak and strong interactions, 
it is assumed that only the oriented knots appear and they 
appear with internal
state functions that are irreducible representations of $SU_q(2)$.  These
state functions will be labelled by the quantum numbers of the standard
theory.  In this way we shall assume that
the quantum knot is defined by 
irreducible representations of its symmetry group $SL_q(2)$.

A similar characterization of a fundamental quantum system by irreducible
representations of its symmetry group is provided by the wave function
of the spherical top, $D^j_{mm^\prime}$, where $D^j_{mm^\prime}$ is an irreducible representation of $O(3)$ and where 
$j(j+1)$
are the eigenvalues of $J^2$, while $m$ and $m^\prime$ are
the eigenvalues of the $z$ components of $\vec J$ in inertial and body
fixed frames.  Another fundamental example is the wave
function of the H atom expressed as the solution of an integral equation
on the group space of $O(3)$, the symmetry group of the hydrogen atom.
There the wave function $D^j_{mm^\prime}$ is an element of an
irreducible representation of $O(3)$; and $2j+1$ is the principal quantum
number, while $m$ and $m^\prime$ are eigenvalues of the $z$ components of
the angular momentum and the Runge-Lenz vector.

In developing the knot model we have begun with the observation that not
only are there 4 trefoils and 4 families of elementary fermions, but also
that there is a unique correspondence between the 4 families and the 4
trefoils, leading to the labelling of the fermionic solitons as the 
following irreducible representations of $SU_q(2)$:  $D^{3t}_{-3t_3-3t_0}$
where $t$ and $t_3$ are isotopic spin and its third component, while
$t_0$ is the hypercharge.  We have shown that these quantum numbers are
connected with the topological characterization of the oriented knot by
\begin{eqnarray}
t &=& \frac{N}{6} \nonumber \\
t_3 &=& -\frac{w}{6} \\
t_0 &=& -\frac{r+1}{6} \nonumber
\end{eqnarray}
where $N,w$, and $r$ are the number of crossings, the writhe, and the
rotation of the knot.$^3$

We now seek a similar representation of the generic knot as the source
of gravitational couplings.

\vskip.5cm

\section{The Generic Knot and the Lorentz Group.}

The quantum algebra, $SL_q(2)$, which describes the symmetry of the generic knot, oriented or not, may be defined as follows.  
Let
\begin{subequations}
\be
L^t_q\epsilon_qL_q = \epsilon_q
\ee
where
\be
\epsilon_q = \left(
\begin{array}{cc}
0 & q_1^{1/2} \\ -q^{1/2} & 0 
\end{array} \right) \qquad q_1 = q^{-1}
\ee
\end{subequations}
Then $L_q$ is a two-dimensional representation of $SL_q(2)$.

If
\be
L_q = \left(
\begin{array}{cc}
a & b \\ c & d
\end{array} \right)
\ee
then
\begin{center}
\begin{tabular}{llll}
$ab= qba$ & $bd = qdb$ & $bc = cb$ & $ad-qbc=1$ \\
$ac=qca$ & $cd=qdc$ & & $da-q_1bc=1$ \hspace{1.25in}(A)
\end{tabular}
\end{center}
There are no finite matrix representations of the elements $(a,b,c,d)$ of
the algebra $(A)$ unless $q$ is a root of unity.

When $q=1$ we have
\begin{subequations}
\begin{eqnarray}
& &L^t\epsilon L = \epsilon \\
& &\epsilon = \left(
\begin{array}{cc}
0 & 1 \\ -1 & 0 
\end{array} \right)
\end{eqnarray}
\end{subequations}
Then $(A)$ is satisfied by complex numbers and (2.3) may be rewritten as
simply
\be
\det~ L = 1
\ee
since
\be
\epsilon_{ij} L_{im}L_{jn} = \epsilon_{mn} \det~ L
\ee
Hence
\be
\lim_{q\to 1} SL_q(2) = SL(2,C)
\ee

We now recall the familiar argument that $SL(2,C)$ describes the Lorentz
group.

Let
\be
P = \vec p\vec\sigma + p_o~1
\ee
Consider
\begin{subequations}
\be
P^\prime =L^+PL
\ee
where
\be
\det~L = 1
\ee
\end{subequations}
Then
\begin{eqnarray}
(P^\prime)^+ &=& P^\prime \\
\det~ P^\prime &=& \det~ P
\end{eqnarray}
By (2.7) and (2.9), Eq. (2.10) may be rewritten as
\be
(p_o^2-\vec p^2)^\prime = p_o^2-\vec p^2 \equiv m_o^2
\ee

Since $L$ has complex matrix elements and $\det~ L=1$ it is a six-parameter
matrix that induces, by (2.8) and (2.11), a Lorentz tranformation on
$(p_o,\vec p)$.

The affine connection of the local Lorentz group leads to the gravitational
interactions.  We shall now postulate that the gravitational couplings
are blind to the orientation of the solitonic knot and  that all
solitons couple to gravity as states of $SL_q(2)$ whether or not the
soliton is oriented.
We shall assume that the action is invariant under global transformation
of the $SL_q(2)$ algebra and in conformity with standard theory, local
transformations of the $SL(2,C)$ algebra.

As the detailed treatment of the gravitational couplings will
follow along the same lines as the electroweak couplings, let us
next describe the algebra that we have used to describe the
oriented knot and the electroweak couplings.

\vskip.5cm

\section{The Oriented Knot and the Electroweak Couplings.}

The quantum algebra $SU_q(2)$, with which we describe the symmetry of
the oriented knot, is the unitary subalgebra of $SL_q(2)$ obtained by
setting
\begin{eqnarray}
d &=& \bar a \\
c &=& -q_1\bar b
\end{eqnarray}
Then $(A)$ reduces to the following
\begin{center}
\begin{tabular}{lll}
$ab= qba$ \qquad & $a\bar a+b\bar b = 1$ 
\qquad & $b\bar b = \bar bb$  \\
$a\bar b =q\bar ba$ \qquad & $\bar aa + q_1^2\bar bb = 1$ 
&  \hspace{1.25in} $(A)^\prime$
\end{tabular}
\end{center}

Denote the $2j+1$-dimensional irreducible representations of $SL_q(2)$
by ${\cal{D}}^j_{mm^\prime}$ and the $2j+1$-dimensional irreducible
representations of $SU_q(2)$ by $D^j_{mm^\prime}$, where
we set in the knot context:
\be
j = \frac{N}{2} \qquad m = \frac{w}{2} \qquad m^\prime = \frac{r+1}{2}
\ee
and $(N,w,r)$ are the number of crossings, the writhe, and the rotation
of the knot.

When the elementary particles are identified as quantum knots, the
topological structure of the knot is determined by the electric charge,
hypercharge, and isotopic spin as stated earlier in Eq. (1.1).$^3$

We again assume that the action is invariant under global transformations
of the $SU_q(2)$ algebra and local transformations of the 
$SU(2)\times U(1)$ group.
The connection of the local $SU(2)\times U(1)$ group leads to the
weak interactions.
The addition of global invariance of the knot symmetries will slightly
modify the standard theory but will not give rise to new fields, as would
be the case if the new symmetries were local.$^2$

The uniform treatment of the 
sources of the gravitational and electroweak
interactions will now be based on the gauge invariance of the knot algebras $SL_q(2)$ and $SU_q(2)$

\vskip.5cm

\section{The Gauge Invariance of $SL_q(2)$ and $SU_q(2)$.}

The $2j+1$-irreducible representations of $SL_q(2)$ and $SU_q(2)$ that we
take to describe the symmetries of the generic knot and the oriented
knot are respectively as follows:

\no \underline{$SL_q(2)$}
\be
{\cal{D}}^j_{mm^\prime} = \sum_{s,t} 
A^j_{mm^\prime}(s,t)
\delta(s+t,n_+^\prime) a^sb^{n_+-s}c^td^{n_--t}
\ee
\no \underline{$SU_q(2)$}
\be
D^j_{mm^\prime} = \sum_{s,t} A^j_{mm^\prime}(s,t) \delta(s+t,n_+^\prime)
(-q_1)^ta^sb^{n_+-s}\bar b^t\bar a^{n_--t}
\ee
where
\be
A^j_{mm^\prime}(s,t) = \left[\frac{\langle n^\prime_+\rangle_1!~
\langle n^\prime_-\rangle_1!}{\langle n_+\rangle_1!~\langle n_-\rangle_1!}
\right]^{1/2}\left\langle\matrix{n_+ \cr s \cr}\right\rangle_1~
\left\langle\matrix{n_- \cr t \cr}\right\rangle_1
\ee
and
\be
\begin{array}{rcl}
n_\pm &=& j\pm m \\ n^\prime_\pm &=& j\pm m^\prime \\
\end{array} \quad
\left\langle\matrix{n \cr s \cr}\right\rangle_1 =
{\langle n\rangle_1!\over \langle s\rangle_1!\langle n-s\rangle_1!}
\quad \langle n\rangle_1 = {q_1^{2n}-1\over q_1^2-1} 
\ee
The algebra $(A)$ of $SL_q(2)$ is invariant under the following gauge
transformation
\be
\begin{array}{rcl}
a^\prime = e^{i\varphi_a}a \qquad & &b^\prime = e^{i\varphi_b}b \\
d^\prime = e^{-i\varphi_a}d \qquad & &c^\prime = e^{-i\varphi_b}c
\end{array}
\ee
Let us examine the resulting transformation of an arbitrary term of (4.1):
\be
(a^{n_a}b^{n_b}c^{n_c}d^{n_d})^\prime = e^{i\varphi_a(n_a-n_d)}
e^{i\varphi_b(n_b-n_c)} (a^{n_a}b^{n_b}c^{n_c}d^{n_d})
\ee
But by (4.1)-(4.4),
\begin{eqnarray}
n_a-n_d = s+t-n_- = n_+^\prime -n_- = m^\prime + m \\
n_b-n_c = n_+-s-t = n_+-n_+^\prime = m-m^\prime
\end{eqnarray}
Then
\be
(a^{n_a}b^{n_b}c^{n_c}d^{n_d})^\prime = e^{i\varphi_a(m+m^\prime)}
e^{i\varphi_b(m-m^\prime)}(a^{n_a}b^{n_b}c^{n_c}d^{n_d})
\ee
Then every term of (4.1) is multiplied by the same factor:
\be
e^{i\varphi_a(m+m^\prime)}e^{i\varphi_b(m-m^\prime)}
\ee
independent of $j$.

Hence the gauge transformation (4.5) of the $SL_q(2)$ algebra
$(A)$ induces the following gauge transformation on the
irreducible representations of $SL_q(2)$
\be
{\cal{D}}^{j~~\prime}_{mm^\prime} = 
e^{i(m+m^\prime)\varphi_a}e^{i(m-m^\prime)\varphi_b}
{\cal{D}}^j_{mm^\prime}
\ee
and in particular one gets back (4.5) when $j=1/2$.

Let us rewrite (4.11) as follows
\be
{\cal{D}}^j_{mm^\prime} = e^{i(\varphi_a+\varphi_b)m}
e^{i(\varphi_a-\varphi_b)m^\prime} {\cal{D}}^j_{mm^\prime}
\ee
All of the preceding relations still hold when one passes from
$SL_q(2)$ to $SU_q(2)$ by setting
\be
\begin{array}{rcl}
d &=& \bar a \\
c &=& -q_1\bar b
\end{array}
\ee
Then (4.12) becomes
\be
D^{j~~\prime}_{mm^\prime} = e^{i\varphi(w)Q(w)}
e^{i\varphi(r)Q(r)}D^j_{mm^\prime}
\ee
where
\begin{eqnarray}
Q(w) &\equiv& km = k \frac{w}{2} \\
Q(r) &\equiv& km^\prime = k \frac{r+1}{2}
\end{eqnarray}

Here we have set $m=w/2$ and $m^\prime = \frac{r+1}{2}$ as in
(3.3).  In this way the (4.12) phase transformation introduces
two knot charges, $Q(w)$ and $Q(r)$, associated with the writhe
and rotation respectively.  These are directly related by
(3.3) and (1.1) to the charges $t_3$ and $t_0$ of the standard
model as shown in Ref. 3.  These relations may also be expressed as
\be
D^j_{mm^\prime} = D^{3t}_{-3t_3-3t_0}
\ee
The $t_3$ and $t_0$ charges stem from the $SU(2)\times U(1)$
symmetries of the standard model while the writhe and rotation
charges originate in the $SU_q(2)$ symmetry of the knot.

\vskip.5cm

\section{Gravitational ``Charges".}

The knot and gravitational symmetries are also closely related
since the former symmetry is characterized by $SL_q(2)$ and the
latter by gauged $SL(2,C)$.  Let us next compare the knot and
gravitational ``charges".

We shall assume that the ${\cal{D}}^{3/2}_{mm^\prime}$ are
those internal wave functions of the fermionic soliton that
interact with the gravitational field just as the 
$D^{3/2}_{mm^\prime}$ are the internal wave functions that
interact with the electroweak fields.  In both cases
$m = \frac{w}{2}$ and $m^\prime = \frac{r+1}{2}$, since the
gravitational and electroweak fields probe different aspects of the single solitonic structure defined by $w$ and $r$.

The irreducible representations ${\cal{D}}^j_{mm^\prime}$ of
$SL_q(2)$ are shown in (4.1) and the gauge transformations on
${\cal{D}}^j_{mm^\prime}$ are given by (4.11).  To pass from
$SL_q(2)$ to $SL(2,C)$ set $q=1$ as we have done in (2.3).  The
algebra $(A)$ is then satisfied by complex numbers and
$\det~L=1$ as in (2.4).  Then the irreducible representations
${\cal{D}}^j_{mm^\prime}(q=1)$ of $SL(2,C)$ may be obtained from
(4.1) as
\be
{\cal{D}}^j_{mm^\prime}(q=1) = \sum_{s,t} 
A^j_{mm^\prime}
(q=1) \delta(s+t,j+m^\prime)\delta(ad-bc,1)
a^sb^{n_+-s}c^td^{n-t}
\ee
where the $a,b,c,d$ are now complex commuting numbers.

If the complex numbers $a,b,c,d$ are subjected to the gauge
transformations (4.5) then the algebra $(A)$ with $q=1$ is
invariant but the ${\cal{D}}^j_{mm^\prime}(q=1)$ are gauge
transformed since the complete set of equations from (4.6) and (4.12) is unchanged and therefore (4.12) reads as follows when
$q=1$:
\be
{\cal{D}}^j_{mm^\prime}(q=1) = 
e^{i(\varphi_a+\varphi_b)m}
e^{i(\varphi_a-\varphi_b)m^\prime}{\cal{D}}^j_{mm^\prime}(q=1)
\ee

Since we are requiring all interactions to be invariant under
the gauge transformations (4.5) that leave the knot algebra
invariant, all interactions are therefore also invariant under
the transformations (5.2).  Then $m$ and $m^\prime$
are conserved in every gravitational interaction and define two 
gravitational ``charges", $Q_1$ and $Q_2$.

The values of the two conserved charges, $Q_1$ and $Q_2$, mass and 
spin, associated with the Lorentz group are 
fixed by specifying their source and hence by
the assignment of $m$ and $m^\prime$.   
Therefore $Q_1$ and $Q_2$ may be regarded as functions of the
conserved $m$ and $m^\prime$:
\begin{eqnarray}
Q_1 &=& f_1(m,m^\prime) \\
Q_2 &=& f_2(m,m^\prime)
\end{eqnarray}
where again $m = \frac{w}{2}$ and $m^\prime = \frac{r+1}{2}$.

Using the factorization of a generic matrix into the product of a
Hermitian and a unitary matrix, we may explicitly display $Q_1$
and $Q_2$ by writing the two-dimensional representation of the
Lorentz group as the product of a boost, $H$, and a unitary
rotation $U$
\be
L = HU
\ee
where $H$ is a Hermitian matrix with $\det~H=1$ and $U$ is a 
$SU(2)$ matrix.  Then we may write
\begin{subequations}
\be
L = \left(
\begin{array}{cc}
\pi_0+\pi_3 & \pi_1-i\pi_2 \\
\pi_1+i\pi_2 & \pi_0-\pi_3 
\end{array} \right)~ e^{\frac{i}{2}\vec\sigma\vec\theta}
\ee
where
\be
\pi_k = p_k/m_0
\ee
\end{subequations}
Here $(\vec p,p_0)$ is the four-momentum of the soliton and
$m_0$ is its rest mass.  Then $m_0H$ is the matrix defined in (2.7), and $L$ is a two-dimensional representation of the 
Lorentz group.  Here $U$ depends on the spin $\vec\sigma/2$
of the fermionic soliton, and $H$
depends on its rest mass $m_0$.  The two ``charges" $Q_1$ and
$Q_2$ in (5.3) and (5.4) may be identified with the rest mass and
any component, say $s_3$ of the spin.  (If one bases this 
discussion on the Poincar\'e group instead of the complex Lorentz group, then the spin is defined by the Pauli-Lubansky
vector and the mass is again defined by $p_\mu$.)  

\vskip.5cm

\section{The Gravitational and Electroweak Interactions.}

The affine connection of the local Lorentz group leads to the
gravitational couplings.  Since we have postulated that the
gravitational couplings are blind as to whether the knot is
oriented, we shall assume
that all solitons enter the gravitational interaction as states
of $SL_q(2)$.  Therefore we shall express the electroweak and
gravitational interactions as follows:
\be
\bar D\bar LWLD + \bar RWR
\ee
and
\be
\tilde{\cal{D}}\bar L\Gamma L{\cal{D}} + R\Gamma R
\ee
where $LD$ and $L{\cal{D}}$ describe left-chiral normal modes with $D$
and ${\cal{D}}$ representing $SU_q(2)$ and $SL_q(2)$ symmetries
respectively, and where $W$ and $\Gamma$ are the electroweak
and gravitational connections respectively.  Here $R$ represents
a right-chiral singlet which is also an internal singlet.
An oriented knot is represented in both
(6.1) and (6.2) and an unoriented knot would appear in only the
gravitational interaction (6.2).  Here $\bar D$ is the adjoint of
$D$ computed according to (4.2) and
\be
\begin{array}{rcl}
a \leftrightarrow \bar a & & \overline{xy} = \bar y~\bar x \\
b \leftrightarrow \bar b & & 
\end{array}
\ee
Likewise $\tilde{\cal{D}}$ is the adjoint of ${\cal{D}}$
computed according to (4.1) and
\be
\begin{array}{rcl}
a &\leftrightarrow& \tilde d \qquad \widetilde{xy} = 
\tilde{y}\tilde{x} \\
b &\leftrightarrow& \tilde c
\end{array}
\ee
Under the gauge transformations (4.11) and (4.14) 
$\tilde{\cal{D}}{\cal{D}}$ and $\bar DD$ are invariant by (4.7)
and (4.8).  Therefore (6.1) and (6.2) are invariant under gauge
transformations (4.11) and (4.14).  

The interactions (6.1) and
(6.2) arise in the covariant derivative terms of the following
form
\be
\bar\psi\nabla\psi
\ee
where
\be
\nabla = \partial + W + \Gamma
\ee
The term (6.5) is invariant under the local group of the standard
theory but also under the global gauge transformations (4.11) and
(4.14).

To evaluate (6.1) and (6.2) we next turn to the spectrum of 
$(SL_q(2)$.

\vskip.5cm

\section{Spectrum of $SL_q(2)$ Algebra $(A)$.}

Since $b$ and $c$ commute, they have common eigenstates.  Let
$|0\rangle$ be designated as a ground state and let
\begin{eqnarray}
b|0\rangle &=& \beta|0\rangle \\
c|0\rangle &=& \gamma|0\rangle \\
bc|0\rangle &=& \beta\gamma|0\rangle
\end{eqnarray}

We assume that $b$ and $c$ are Hermitian:
\begin{eqnarray}
b &=& \bar b \\
c &=& \bar c
\end{eqnarray}
Then the eigenvalues $\beta,\gamma$ are real and the
eigenfunctions are orthogonal.

From the algebra we see that
\be
bc|n\rangle = E_n|n\rangle
\ee
where
\begin{eqnarray}
|n\rangle &\sim& d^n|0\rangle \\
E_n &=& q^{2n}\beta\gamma
\end{eqnarray}
Here $d$ and $a$ are raising and lowering operators respectively.
\begin{eqnarray}
d|n\rangle &=& \lambda_n|n+1\rangle \\
a|n\rangle &=& \mu_n|n-1\rangle
\end{eqnarray}
Then
\begin{eqnarray}
ad|n\rangle &=& a\lambda_n|n+1\rangle \nonumber \\
&=& \lambda_n \mu_{n+1}|n\rangle \\
da|n\rangle &=& d \mu_n|n-1\rangle \nonumber \\
&=& \mu_n \lambda_{n-1}|n\rangle
\end{eqnarray}
From the algebra $(A)$
\begin{eqnarray}
(1+qbc)|n\rangle &=& \lambda_n \mu_{n+1}|n\rangle \\
(1+q_1bc)|n\rangle &=& \mu_n\lambda_{n-1}|n\rangle
\end{eqnarray}
If there is a highest state $M$, and a lowest state $M^\prime$,
then
\be
\lambda_M = \mu_{M^\prime} = 0 \qquad M^\prime < M
\ee
By (7.13) and (7.14)
\begin{eqnarray}
& &(1+qbc)|M\rangle = 0 \\
& &(1+q_1bc)|M^\prime\rangle = 0
\end{eqnarray}
Then by (7.8)
\be
q^{2M+1}\beta\gamma = q^{2M^\prime-1}\beta\gamma
\ee
or
\be
(q^2)^{M-M^\prime+1} = 1
\ee
As we continue to assume that $q$ is real,
\be
M^\prime = M+1
\ee
Since (7.15) and (7.20) are contradictory, there may be either a
highest or a lowest state, but not both.

According to the general rules of our model,$^{1,2,3}$ the
individual states of excitation of the soliton represented by
$D^j_{mm^\prime}$ are $D^j_{mm^\prime}|n\rangle$.  Since the empirical evidence appears to restrict the number of states, 
there must be an externally required
physical boundary condition to cut off the otherwise infinite
spectrum that is formally required by (7.15) and (7.20).

\vskip.5cm

\section{Evaluation of Electroweak and Gravitational Interactions.}

To evaluate the interactions (6.1) and (6.2) between 
left-chiral solitonic
states let us first consider the generic electroweak interaction:
\be
\bar{\cal{F}}_3{\cal{W}}_2{\cal{F}}_1
\ee
where
\begin{eqnarray}
{\cal{F}}_1 &=& F_1(p,s,t) D^{3/2}_{m_1m_1^\prime}|n_1\rangle
\\
{\cal{F}}_3 &=& F_3(p,s,t) D^{3/2}_{m_3m_3^\prime}|n_3\rangle
\\
{\cal{W}}_2 &=& W_2(p,s,t) D^j_{m_2m_2^\prime}
\end{eqnarray}
Here $F(p,w,t)$ and $W(p,s,t)$ are the standard 
left chiral fermionic and
electroweak normal modes where $(p,s,t)$ represent momentum,
spin and isotopic spin.  The $D^j_{mm^\prime}$ are the internal
state factors attached to the standard normal modes of the 
electroweak connection.

Then (8.1) may be rewritten as
\be
(\bar F_3W_2F_1)\langle n_3|\overline{D^{3/2}_{m_3m_3^\prime}}
D^j_{m_2m_2^\prime}D^{3/2}_{m_1m_1^\prime}|n_1\rangle
\ee
The correction to the standard matrix element appears in the
second factor:
\be
\langle n_3|\bar D^{3/2}_{m_3m_3^\prime}D^j_{m_2m_2^\prime}
D^{3/2}_{m_1m_1^\prime}|n_1\rangle
\ee
We require that this internal factor or form factor be invariant
under $U_a(1)\times U_b(1)$ defined by (4.5) and (4.14).  It
then follows that$^3$
\begin{eqnarray}
m_1+m_2 &=& m_3 \\
m_1^\prime + m_2^\prime &=& m_3^\prime
\end{eqnarray}
i.e., the topological and isotopic observables whose eigenvalues
are $m$ and $m^\prime$ are conserved.

The conservation of $m$ and $m^\prime$ then make it possible to
label the $D^j_{mm^\prime}$ by either topological or isotopic
quantum numbers as follows:
\begin{eqnarray}
j &=& 3t \qquad m = -3t_3 \qquad m^\prime = -3t_0 \\
j &=& \frac{N}{2} \quad~~ m = \frac{w}{2} \qquad~~ m^\prime
= \frac{r+1}{2}
\end{eqnarray}
and to follow the same rules for the bosons as for the fermions.
These identifications are discussed in Refs. 2 and 3.

We shall assume the same construction for the gravitational matrix
element corresponding to (8.6) but also in accordance
with (6.2), namely:
\be
\langle n_3|\tilde{\cal{D}}^{3/2}_{m_3m_3^\prime}
{\cal{D}}^j_{m_2m_2^\prime} {\cal{D}}^{3/2}_{m_1m_1^\prime}
|n_1\rangle
\ee
where the ${\cal{D}}^{3/2}_{mm^\prime}$ correspond to 
$SL_q(2)$ rather than
$SU_q(2)$ and where $\tilde{\cal{D}}^{3/2}_{mm^\prime}$ 
is the modified
adjoint computed according to (4.1) and (6.4).

We now propose to use the same conservation laws (8.7) and 
(8.8) to assign quantum
numbers to the internal boson factor 
${\cal{D}}^{j_2}_{m_2m_2^\prime}$
in (8.11).  For the graviton, however, we choose $j=0$.  
Therefore we set
\be
{\cal{D}}^j_{m_2m_2^\prime} = {\cal{D}}^0_{00} = 1
\ee
Then the graviton is an unknotted clockwise loop.
Therefore the correction factor or form factor,
in the gravitational case is
simply
\be
\langle n_3|\tilde{\cal{D}}^{3/2}_{m_3m_3^\prime}
{\cal{D}}^{3/2}_{m_3m_3^\prime}|n_1\rangle \delta(n_3,n_1)
\ee
In this section the distinction that we have made between the
electroweak and gravitational form factors 
lies entirely in the
use of ${\cal{D}}$ and $D$, while the labelling (quantum numbers)
is the same for both
${\cal{D}}$ and $D$ since they describe different aspects of the same trefoil.

\vskip.5cm

\section{The Quark-Gluon Interaction.}

Let us consider the tensor products
\be
\begin{array}{rcl}
a^\prime &=& U_aa \qquad~~ b^\prime = U_bb \\
d^\prime &=& U^{-1}_ad \qquad c^\prime = U_b^{-1}c
\end{array}
\ee
where $U_a$ and $U_b$ are commuting elements of a unimodular
unitary group that also commute with $(a,b,c,d)$.  Then
$(a^\prime,b^\prime,c^\prime,d^\prime)$ satisfy the same
algebra $(A)$ as $(a,b,c,d)$.

Let ${\cal{D}}^j_{mm^\prime}$ be an element of an irreducible
unitary representation of the algebra $(A)$.  Then by the same
argument that leads to (4.11)
\be
{\cal{D}}^{j~~\prime}_{mm^\prime} = U_a^{m+m^\prime}
U_b^{m-m^\prime}{\cal{D}}^j_{mm^\prime}
\ee
set
\begin{eqnarray}
U_a &=& e^{iQ_a\theta_a} \qquad {\rm Tr}~Q_a = 0 \\
U_b &=& e^{iQ_b\theta_b} \qquad {\rm Tr}~Q_b = 0
\end{eqnarray}
where $Q_a$ and $Q_b$ are both Hermitian and diagonal and hence
real.  Let us now restrict $U_a$ and $U_b$ to the center of the
$SU(3)$ group that describes the strong couplings.  These
couplings, the gluon-quark interactions in the standard theory,
have the following form
\begin {eqnarray}
& &{\cal{L}}_{\rm int}(\rm {quark-gluon}) \nonumber \\
& &\mbox{}= -\frac{g_0}{2}\bar q(Q_AA\!\!\!/ + 
Q_BB\!\!\!/)q-\frac{g_0}{\sqrt{2}}
[\bar q(\tau_{21}
X\!\!\!/+\tau_{31}Y\!\!\!/ + \tau_{32}Z\!\!\!/)q+{\rm c.c.}]
\end{eqnarray}
where the generators of $SU(3)$ in the fundamental representation
are
\be
Q_A = \left(
\begin{array}{ccc}
\frac{1}{2} & & \\ & -\frac{1}{2} & \\ & & 0
\end{array} \right) \qquad
Q_B = \frac{1}{\sqrt{3}} \left(
\begin{array}{ccc}
\frac{1}{2} & & \\ & \frac{1}{2} & \\ & & -1
\end{array} \right)
\ee
and
\be
\tau_{12} = \left(
\begin{array}{ccc}
0 & 1 & 0 \\ 0 & 0 & 0 \\ 0 & 0 & 0
\end{array} \right) \qquad \tau_{13} = \left(
\begin{array}{ccc}
0 & 0 & 1 \\ 0 & 0 & 0 \\ 0 & 0 & 0
\end{array} \right) \qquad \tau_{23} = \left(
\begin{array}{ccc}
0 & 0 & 0 \\ 0 & 0 & 1 \\ 0 & 0 & 0
\end{array} \right)
\ee

In the present model the quarks have a knot structure as already
described.  If a knot structure is also assigned to the gluons,
then each term of (9.5) will acquire a knot factor.  We shall
require that the so amended action be invariant not only under
local $SU(3)$ strong, as required by standard theory, but also
under the global transformation (9.1) and therefore (9.2).  These
global transformations may be re-expressed as
\begin{eqnarray}
{\cal{D}}^j_{mm^\prime} &=& (U_aU_b)^m (U_aU_b^{-1})^{m^\prime}
{\cal{D}}^j_{mm^\prime} \nonumber\\
&=& e^{i(Q_a\theta_a+Q_b\theta_b)m}
e^{i(Q_a\theta_a-Q_b\theta_b)m^\prime}{\cal{D}}^j_{mm^\prime} 
\nonumber\\
&=& e^{iQ_A\theta_A}e^{iQ_B\theta_B}{\cal{D}}^j_{mm^\prime}
\end{eqnarray}
where
\begin{subequations}
\begin{eqnarray}
Q_A\theta_A &=& (Q_a\theta_a+Q_b\theta_b)m \\
&=& (Q_a\theta_a + Q_b\theta_b)(-3t_3) 
\end{eqnarray}
\end{subequations}
where
\begin{subequations}
\begin{eqnarray}
Q_B\theta_B &=& (Q_a\theta_a-Q_b\theta_b)m^\prime \\
&=& (Q_a\theta_a-Q_b\theta_b)(-3t_0)
\end{eqnarray}
\end{subequations}
The same equations (9.8)-(9.10) hold if ${\cal{D}}$ is
replaced by $D$.
We shall now identify $Q_A$ and $Q_B$ as the color charge and
the color hypercharge respectively as described by the standard
theory.  Here $Q_A \sim t_3$ and $Q_B \sim t_0$ according
to Eqs. (9.9b) and (9.10b).  (Note, however, that $Q_A\sim
t_3$ and not $t_3+t_0$.)  The color charge and hypercharge are
given by (9.6) and Table 1.

\begin{center}
{\bf Table 1.}
\end{center}
\[
\begin{array}{rcc}
& Q_A~(\mbox{color charge}) \qquad& Q_B~(\mbox{color hypercharge})
\\
\mbox{red}~~ & 1/2 \qquad& \sqrt{3}/6 \\
\mbox{yellow}~~ & -1/2 \qquad& \sqrt{3}/6 \\
\mbox{green}~~ & 0 \qquad& -\sqrt{3}/3
\end{array}
\]
These are the color charge and color hypercharge of the fermionic
soliton with $r=-2$ (or weak hypercharge =1/6).  If $r=2$ (or
$t_0=-1/2$) $Q_A=Q_B=0$.

In Table 2 one finds for the solitons of the $u$-family and the
$d$-family the explicit relations between the weak and the
strong charges that are shown in Table 1.  There the strong color
charge $Q_A$ is related to the weak $t_3$ and the strong color
hypercharge $Q_b$ is related to $t_0$ as indicated in (9.9b) and
(9.10b).
\begin{center}
{\bf Table 2.}
\end{center}
\[
\begin{array}{c|ccccccccc}
& t_3 & t_0 & D^{3/2}_{-3t_3-3t_0} & Q_{AR} & Q_{AY} & Q_{AG} &
Q_{BR} & Q_{BY} & Q_{BG} \\
\hline
u & \frac{1}{2} & \frac{1}{6} & D^{3/2}_{-3/2~-1/2} & t_3 &
-t_3 & 0 & \sqrt{3}t_0 & \sqrt{3}t_0 & -2\sqrt{3}t_0 \\
d & -\frac{1}{2} & \frac{1}{6} & D^{3/2}_{3/2~-1/2} & -t_3 &
t_3 & 0 & \sqrt{3}t_0 & \sqrt{3}t_0 & -2\sqrt{3}t_0 \\
\end{array}
\]
As illustrated in Table 1 there are three varieties of solitons
with $r=-2$ that differ in gluon charge and are conventionally
labelled $R,Y$, and $G$.  There is only one variety of soliton
with $r=2$.  There are eight gluons including the charged
triplet $(X,Y,Z)$ and their antigluons as well as the uncharged
$A$ and $B$ displayed in Eq. (9.5).  The charges of $(X,Y,Z)$
are fixed by the conservation of gluon charge and hypercharge
together with the assignments to $u$ and $d$ of $Q_A$ and
$Q_B$ shown in Table 2.  The charges of $(X,Y,Z)$ are shown in
Table 3 expressed in terms of $t_3$ and $t_0$ for the $u$
soliton.

\begin{center}
{\bf Table 3.}
\end{center}
\[
\begin{array}{c|cc}
& Q_A \qquad~~~& Q_B \\ \hline
X ~~~&-2t_3 \qquad~~~& 0 \\
Y ~~&-t_3 \qquad& -3\sqrt{3}t_0 \\
Z ~~~&t_3 \qquad~~~& -3\sqrt{3}t_0 \\
\end{array}
\]
The strong charges that are, according to Eqs. (9.9b) and (9.10b)
and to Tables 1 and 3, simply related to the weak $t_3$ and
$t_0$, are in turn simply related to the topological description
of the knot source since $t_3 = -\frac{w}{6}$ and
$t_0 = -\frac{r+1}{6}$.

The content of the preceding tables may be simply expressed by
the following relations (induced by the transformations (9.1)
on the knot algebra).

By (9.9b) and (9.10b)
\begin{eqnarray}
Q_A &=& \stackrel{\circ}{Q}_At_3 \\
Q_B &=& \stackrel{\circ}{Q}_Bt_0
\end{eqnarray}
where
\be
{\rm Tr}~ \stackrel{\circ}{Q}_A = {\rm Tr}~\stackrel
{\circ}{Q}_B = 0
\ee
and
\be
t_3 = \pm \frac{1}{2} \qquad t_0 = \frac{1}{6}
\ee
Since (9.11) and (9.12) must agree with (9.6) of the standard
theory, we have
\begin{eqnarray}
\stackrel{\circ}{Q}_A &=& \left(
\begin{array}{ccc}
\pm 1 & & \\ & \mp 1 & \\ & & 0
\end{array}\right) \\
\stackrel{\circ}{Q}_B &=& \sqrt{3} \left(
\begin{array}{ccc}
1 & & \\ & 1 & \\ & & -2 
\end{array} \right)
\end{eqnarray}
where the upper and lower signs correspond to $u$ and $d$
respectively.
The linear relation between $(Q_A,Q_B)$ and $(w,r)$ that is
implied by (9.11) and (9.12) is completed by (9.15) and (9.16),
as well as by $t_3 = \frac{w}{2}$ and $t_0 = \frac{1}{2} (r+1)$.

We shall now describe a modification of (9.5) that is similar
to (8.11) and (8.6) where we have introduced knot form factors
for the gravitational and electroweak interactions.  For the
gluon form factors we asume that the quark knot is oriented 
and therefore choose 
\be
\langle n_3|\bar D^{3/2}_{m_3m_3^\prime}(c_3)
D^j_{m_2m_2^\prime}(c_2)
D^{3/2}_{m_1m_1^\prime}(c_1)|n_1\rangle
\ee
where $m=w/2$ and $m^\prime = \frac{r+1}{2}$ as in the
gravitational and electroweak cases.  Here the first and third
factors are the internal state functions of the interacting
quarks while the second factor is the internal state function
of a gluon.  Each quark and gluon may acquire one of three
colors according to (9.11) and (9.12) but the internal state
function does not depend explicitly on the color index $(c)$.
The transformation of $D^j_{mm^\prime}(c)$ under $U_a(1)\times
U_b(1)$ does depend on $c$ according to (9.8) holding for
$D$.
In addition to the conservation of color in the standard theory
there is now the conservation of $m$ and $m^\prime$ 
that stems from invariance under global $U_a(1)\times U_b(1)$ and
fixes $m_2$ and $m_2^\prime$ in terms of the $m$ and 
$m^\prime$ of the initial and final quarks.

The different facets of the knotted soliton required to describe
its different couplings may be related by the charges
(integrals of the motion) corresponding to these different
couplings.  First, as a topological object the knotted soliton
has two ``charges", the writhe and the rotation.  Next, as a
source of the electroweak fields it also has two charges, the
electric charge and the hypercharge coming from $SU(2)\times
U(1)$.  Then there are again two integrals of the motion, mass
and spin, stemming from the Lorentz group, which, as a local
group, leads to the gravitational couplings.  Finally, as a 
source of the gluon field, the quarks have two additional
charges, color charge and color hypercharge, eigenvalues of the
commuting generators of $SU(3)$.  In summary, if one labels the
particle by assigning the conserved $w$ and $r$, then, depending
on the context, two conserved charges $Q_A$ and $Q_B$ are
fixed by
\be
\begin{array}{rcl}
Q_A &=& F_A(w,r) \\
Q_B &=& F_B(w,r)
\end{array}
\ee 
where $Q_A$ and $Q_B$ come from one of the groups $SL(2c)$,
$SU(2)\times U(1)$, or $SU(3)$ and correspond to the different
mappings of $(w,r)$ onto $(Q_A,Q_B)$ that are denoted by
$F_A$ and $F_B$.

In the case of the electroweak and the gluon couplings the
relations $F_A$ and $F_B$ are linear.  In the Lorentz or
gravitational case one may write $s_3 = \pm\frac{w}{6}$ for
the spin, but the relation of the mass to $(w,r)$ is not linear
and will be considered next.

\vskip.5cm

\section{Hamiltonian and Integrals of the Motion of the Quantum
Knot.}

Let the Hamiltonian of the quantum knot be $H(b,c)$.  Then
since the generic knot is defined by (4.1), we have
\begin{eqnarray}
H(b,c){\cal{D}}^j_{mm^\prime}|n\rangle &=& H(b,c)\left[\sum_{s,t}
A^j_{mm^\prime}\delta(s+t,n^\prime_+)a^sb^{n_+-3}c^t
d^{n_--t}\right]|n\rangle \\
&=& {\cal{D}}^j_{mm^\prime}H(q_1^{n_a-n_d}b,q_1^{n_a-n_d}c)
|n\rangle
\end{eqnarray}
where $n_a$ and $n_d$ are the exponents of $a$ and $d$ 
respectively.  Then by relations like (7.8)
\be
\begin{array}{rcl}
H(b,c){\cal{D}}^j_{mm^\prime}|n\rangle &=& {\cal{D}}^j_{mm^\prime}H(q_1^{n_a-n_d}q^n\beta,q_1^{n_a-n_d}q^n\gamma|n\rangle\\
&=& E^j_{mm^\prime}(n){\cal{D}}^j_{mm^\prime}|n\rangle
\end{array}
\ee
where
\begin{subequations}
\be
E^j_{mm^\prime}(n) = H(\lambda\beta,\lambda\gamma)
\ee
and
\be
\lambda = q^{n-(m+m^\prime)}
\ee
\end{subequations}
by (4.7).  Therefore the  ${\cal{D}}^j_{mm^\prime}$ are 
eigenstates of $H(b,c)$
and the indices on ${\cal{D}}^j_{mm^\prime}$ are the
eigenvalues of the integrals of the motion.  The eigenvalues of
$H(b,c)$ are $H(\lambda\beta,\lambda\gamma)$ by (10.4).

The operators that represent integrals of the motion may be
expressed in terms of an elementary operator $\omega_x$ that
may be defined by its action on every term of ${\cal{D}}^j_{mm^\prime}$ as follows:
\be
\omega_x(\ldots x^{n_x}\ldots) = n_x(\ldots x^{n_x}\ldots) 
\qquad x = (a,b,c,d)
\ee
i.e., $\omega_x$ acts like $x\frac{\partial}{\partial x}$.

Then define
\begin{eqnarray}
{\cal{J}} &=&\frac{1}{2} (\omega_a + \omega_b + \omega_c +
\omega_d) \\
{\cal{W}} &=& \frac{1}{2} (\omega_a - \omega_d + \omega_b -
\omega_c) \\
{\cal{R}} &=& \frac{1}{2} (\omega_a - \omega_d - \omega_b +
\omega_c)
\end{eqnarray}
When ${\cal{J}}$, ${\cal{W}}$, and ${\cal{R}}$ act on
${\cal{D}}^j_{mm^\prime}$ one finds by (10.5)
\begin{eqnarray}
{\cal{J}}~{\cal{D}}^j_{mm^\prime} &=& j~{\cal{D}}^j_{mm^\prime}
\\
{\cal{W}}~{\cal{D}}^j_{mm^\prime} &=& m~{\cal{D}}^j_{mm^\prime}
\\
{\cal{R}}~{\cal{D}}^j_{mm^\prime} &=& m^\prime~
{\cal{D}}^j_{mm^\prime}
\end{eqnarray}
The operators $({\cal{J}},{\cal{W}},{\cal{R}})$ are all
interpreted as integrals of the motion with different physical
identifications of the eigenvalues $m$ and $m^\prime$ for
$SL_q(2)$, $SU_q(2)$, and $SU(2)$ in the different examples we
have discussed.  For $SL_q(2)$ $m$ and $m^\prime$ will be
functions of mass and spin; for $SU_q(2)$ $m$ and $m^\prime$
will be functions of charge and hypercharge or alternatively of
writhe and rotation.  For $SU(2)$ one has the non-solitonic
examples of the spherical top and $H$-atom mentioned in the
introduction.

We also introduce the inversion operator transforming ${\cal{D}}$
into $\tilde{\cal{D}}$:
\be
\tilde{\cal{D}} = {\cal{I}}{\cal{D}}
\ee
${\cal{I}}$ which is defined by (6.4) 
interchanges the exponents of $a$ and $d$ and 
of $b$ and $c$ and transforms the algebra
$A$ into the same algbra with $q\to q_1$.  Note that
\be
\begin{array}{rcl}
{\cal{I}}~{\cal{J}} &=& {\cal{J}} \\
{\cal{I}}~{\cal{W}} &=& -{\cal{W}} \\
{\cal{I}}~{\cal{R}} &=& -{\cal{R}}
\end{array}
\ee
In this section the state function of the knot has been taken
to be ${\cal{D}}^j_{mm^\prime}$.  All of the statements of this
section remain correct for $D^j_{mm^\prime}$ when one passes
to the subalgebra $(A)^\prime$.

\vskip.5cm

\section{Masses of the Trefoils.}

In an atomic or nuclear system the energy levels are calculated
with the aid of a Hamiltonian which in turn is determined by
the assumed potential energy.  If on the other hand the system
is defined, as it is here, solely by its symmetry group, the
only restriction on the Hamiltonian is provided by the group
itself.  We have taken the Hamiltonian of the knot to be a
function of the operators $b$ and $c$, say, $f(b,c)$.  Since
the knot is here defined by state functions 
${\cal{D}}^{3/2}_{\frac{w}{2}\frac{r+1}{2}}$ we see by (10.4)
that the eigenvalues of $f(b,c)$ are given by
\begin{eqnarray}
E(w,r,n) &=& f(\lambda\beta,\lambda\gamma) \\
\lambda &=& q^{n-(w+r+1)/2}
\end{eqnarray}
where $E(w,r,n)$ will be interpreted as the mass of the n$^{\rm th}$ excited state of the trefoil with writhe $w$ and rotation 
$r$.

The choice of the function $f(b,c)$ can be fixed by the
requirement that $f(b,c)$ agree with the mass term in the
action of the standard theory, namely
\be
\bar L\varphi R + \bar R\bar\varphi L
\ee
In the standard theory $L$ and $R$ are the left- and right-chiral
components of the fermion field and $\varphi$ is the Higgs
scalar.  $L$ and $\varphi$ are $SU(2)$ isotopic doublets and $R$ is an isotopic singlet so that $\bar L\varphi R$ is an
isotopic invariant.

In the present model, we assume that $R$ is a singlet in the
knot algebra as well as in the isotopic group.  Then (11.3) 
reduces to
\be
\bar L\varphi + \bar\varphi L
\ee
where we have dropped the external spacetime factors in
$L,\varphi$ and $R$ and retained only the internal state
factors.

To translate (11.3) into knot language, note that an oriented
knot is represented in gravitational interactions by
${\cal{D}}^{3/2}_{\frac{w}{2}\frac{r+1}{2}}$ and in electroweak
interactions by $D^{3/2}_{\frac{w}{2}\frac{r+1}{2}}$
according to (6.1) and (6.2).  In the mass term (11.4) we
choose to represent $L$ by the form, ${\cal{D}}^{3/2}_{\frac{w}{2}\frac{r+1}{2}}$, appropriate to gravitational interactions,
since the gravitational field of a particle measures the mass
of that particle.  Since we require that (11.4) be a knot
invariant as well as an isotopic invariant, we finally assume
that $\varphi$ is also represented by 
${\cal{D}}^{3/2}_{\frac{w}{2}\frac{r+1}{2}}$.

Then we fix $f(b,c)$, the Hamiltonian of the knot by setting
\be
f(b,c) \sim \bar L\varphi + \bar\varphi L
\ee
or
\be
f(b,c) \sim\tilde{\cal{D}}^{3/2}_{\frac{w}{2}\frac{r+1}{2}}
{\cal{D}}^{3/2}_{\frac{w}{2}\frac{r+1}{2}}
\ee
By the algebra $(A)$ every expression (11.6) is a function of
$b$ and $c$ only.  Then the masses are
\be
m(w,r,n)\sim\langle n|
\tilde{\cal{D}}^{3/2}_{\frac{w}{2}\frac{r+1}{2}}
{\cal{D}}^{3/2}_{\frac{w}{2}\frac{r+1}{2}}|n\rangle
\ee
Since this expression (11.7) is the same as (8.13), it follows
from this result that the gravitational couplings are proportional to the mass in this model.

In previous work$^{1,2}$ we have compared the empirical masses
of the elementary fermions with
\be
m(w,r,n) \sim \langle n|\bar D^{3/2}_{\frac{w}{2}\frac{r+1}{2}}
D^{3/2}_{\frac{w}{2}\frac{r+1}{2}}|n\rangle
\ee
It is possible to fit the observed masses with either (11.7) or
(11.8) by adjusting $q$ and $\beta$ but the spectrum predicted
by both must be cut off since there appears to be not more
than three members in each family of fermions.  Here, as in our 
earlier work, we assume that the three members of each family 
occupy the three lowest states of energy.

\vskip.5cm

\no {\bf Remarks.}

In assuming that the elementary particles are knots of field
one assigns a unifying role to the knot algebra.  In this way the
electroweak, color, and gravitational symmetries may be
related to the symmetries of the knot.  As a result, charge
and hypercharge, color charge and color hypercharge, as well as
mass and spin of the elementary fermions may be expressed as
functions of the writhe and rotation of the corresponding
trefoils.  These functions are determined jointly by the
gauge invariance of the knot algebra and by the $SU(2)\times
U(1)$, and $SU(3)$, and $SU(2c)$ groups.  In the field theoretic
implementation of this correspondence it is proposed that the
state functions of the quantum knots be attached to the normal
modes of the standard theory with the result that the interaction
terms of the standard theory are multiplied by form factors
reflecting the trefoil origins of the quantum numbers of the
standard theory.

If it is correct that the trefoils play a fundamental role in
the structure of the elementary particles, then it is natural to
ask if the elementary trefoil can be realized as an elementary
boson as well as an elementary fermion, i.e., as a scalar as 
well as a chiral spinor.  (The fact that there are 4 trefoils and
4 classes of fermions as well as a unique correspondence between
them, makes it possible to interpret the 3 members of each
fermionic class as eigenstates of the corresponding knot.)  In
comparing the conjectured fermionic trefoils with the physical
fermions, one finds that the topology of the trefoil, as
defined by the writhe and rotation, correlates uniquely via the
knot algebra with the isotopic structure of the physical fermion
as defined by its charge and hypercharge.  If one conjectures 
that the topologyy in fact determines the charge structure, then
if the scalar trefoil in fact exists, one may also conjecture 
that this trefoil, sharing the topology of the spinor trefoil,
will also share the same charge and hypercharge structure.
This extended model would suggest a search for the charged
scalars with the same charge structure as the fermions.

\vskip.5cm

\no {\bf References.}

\begin{enumerate}
\item R. J. Finkelstein, Int. J. Mod. Phys. A{\bf 20}, 6481
(2005).
\item A. C. Cadavid and R. J. Finkelstein, {\it ibid.} 
A{\bf 21}, 4214 (2006).
\item R. J. Finkelstein, {\it ibid.} A{\bf 22}, 4467 (2007). 
\item L. H. Kauffmann, {\it ibid.} A{\bf 5}, 93 (1990).
\end{enumerate}

\end{document}